\documentclass[pdflatex,sn-mathphys-num,iicol]{sn-jnl}

\usepackage{graphicx}%
\usepackage{multirow}%
\usepackage{amsmath,amssymb,amsfonts}%
\usepackage{amsthm}%
\usepackage{mathrsfs}%
\usepackage[title]{appendix}%
\usepackage{xcolor}%
\usepackage{textcomp}%
\usepackage{manyfoot}%
\usepackage{booktabs}%
\usepackage{algpseudocode}%
\usepackage{listings}%
\UseRawInputEncoding
\usepackage[ruled,vlined,linesnumbered]{algorithm2e}

\theoremstyle{thmstyleone}%
\theoremstyle{thmstyletwo}%

\theoremstyle{thmstylethree}%

\begin{document}


\title{Tunable Nonlocal $ZZ$ Interaction for Remote Controlled-Z Gates Between Distributed Fixed-Frequency Qubits}

\author[1]{\fnm{Benzheng} \sur{Yuan}}

\author[1]{\fnm{Chaojie} \sur{Zhang}}

\author[1]{\fnm{Haoran} \sur{He}}

\author[1]{\fnm{Yangyang} \sur{Fei}}
\author[1]{\fnm{Chuanbing} \sur{Han}}
\author[1]{\fnm{Shuya} \sur{Wang}}
\author[1]{\fnm{Huihui} \sur{Sun}}
\author[1]{\fnm{Qing} \sur{Mu}}
\author[1]{\fnm{Bo} \sur{Zhao}}
\author[1]{\fnm{Fudong} \sur{Liu}}
\author*[1]{\fnm{Weilong} \sur{Wang}}\email{wangwl19888@163.com}
\author*[1]{\fnm{Zheng} \sur{Shan}}\email{shanzhengzz@163.com}

\affil[1]{\orgdiv{Laboratory for Advanced Computing and Intelligence Engineering}, \orgname{Information Engineering University}, \orgaddress{\street{No. 62 Science Avenue}, \city{Zhengzhou}, \postcode{450001}, \state{Henan}, \country{China}}}


\abstract{
Scaling superconducting quantum processors toward fault-tolerant operation will likely require architectures that extend beyond monolithic chips. Modular processors connected by low-loss superconducting links provide a promising route, but implementing entangling gates between remote fixed-frequency qubits remains challenging. Here we propose a distributed architecture in which two synchronously controlled double-transmon couplers mediate the interaction between fixed-frequency transmons in separate packages connected by a 25-cm coaxial cable. The scheme activates a tunable nonlocal $ZZ$ interaction on demand while suppressing residual static coupling, allowing the superconducting link to function as a gate-native interconnect rather than solely as a state-transfer channel. Circuit-level simulations show an on/off ratio exceeding $10^6$ and a remote controlled-Z gate with a projected coherent fidelity of $99.99\%$ under experimentally relevant parameters. Open-system simulations further indicate that, within the representative Markovian noise model considered here, endpoint-qubit decoherence is the largest contribution to gate infidelity, while photon loss in the retained cable modes remains smaller but non-negligible. These results identify DTC-mediated tunable nonlocal coupling as a promising gate primitive for modular superconducting processors based on fixed-frequency qubits.
}





\maketitle

\section{Introduction}\label{sec1}
Fault-tolerant quantum computing (FTQC) will require large numbers of physical qubits operated at fidelities compatible with quantum error correction (QEC) \cite{RN258,RN259,RN260,RN261,RN262,RN263,RN257}. Superconducting circuits have made substantial progress toward this goal through increasingly large monolithic processors \cite{RN166,RN257,zuchongzhi,RN295,RN296}. However, further scaling of a single chip is constrained by packaging complexity, thermal load, and parasitic electromagnetic crosstalk \cite{RN264,RN265}. These constraints have motivated modular superconducting architectures, in which spatially separated quantum processing units (QPUs) are interconnected to operate as a larger distributed processor.

A central requirement for such architectures is the ability to perform entangling operations across module boundaries. Optical platforms have enabled remarkable progress in long-distance quantum communication and networking \cite{pan,su,zhoulai,zongquan,Caleffi,jessica}, but distributed superconducting processors face a distinct hardware requirement. In this setting, inter-chip operations must be fast, deterministic and compatible with repeated QEC cycles. Near-term multichip modules within a single package have demonstrated promising inter-chip coupling and gate operations \cite{RN266,RN267,RN268}, yet their scalability remains constrained by dense wiring and parasitic intra-package crosstalk \cite{RN270}. Macroscopic low-loss superconducting links provide an alternative route for connecting separated modules \cite{RN271,RN272,RN273,RN274,RN275,RN276,RN277}, but converting such links into gate-compatible interconnects remains challenging.

Most approaches to crossing the inter-node boundary have relied on quantum-state transfer \cite{RN291}. These protocols are essential for quantum communication, but using them as the basis for deterministic two-qubit gates introduces additional steps for emission, capture, synchronization and calibration \cite{RN272,RN277}. This overhead is not naturally aligned with the rapid, repeated entangling operations required in QEC cycles. Recent work has therefore explored more direct remote-gate mechanisms. Tunable-coupler-based gmon approaches can provide strong interactions, but often perturb computational-qubit frequencies during switching \cite{RN278,RN279,RN280}. Fixed-frequency approaches based on cross-resonance driving \cite{RN281,RN282} or frequency-tuning-induced $ZZ$ interactions \cite{RN284} can reduce some control demands, but still face trade-offs among frequency crowding, parasitic crosstalk and gate contrast. A remaining challenge is to engineer a remote entangling interaction that is switchable, gate-native and compatible with fixed-frequency qubits.

Here we introduce a distributed superconducting architecture in which two synchronously controlled double-transmon couplers convert a multimode superconducting cable into a switchable nonlocal $ZZ$ gate element. In this scheme, two fixed-frequency transmons in separate packages are each coupled to a double-transmon coupler (DTC) \cite{RN285,RN287,RN288,RN289,RN290}, and the two couplers are connected by a 25-cm coaxial cable. Flux control of the DTCs induces local qubit--cable cross-Kerr interactions, while synchronous control of the two couplers activates an effective nonlocal $ZZ$ interaction between the remote qubits. The cable thereby functions as part of a gate-native interacting system rather than solely as a state-transfer channel. Circuit-level simulations show that the nonlocal interaction can be switched from an idle value below $10^{-5}~{\rm MHz}$ to the megahertz scale, supporting a remote CZ gate with a projected coherent fidelity of $99.99\%$. Open-system simulations further indicate that, within the representative Markovian noise model considered here, endpoint-qubit decoherence gives the largest contribution to gate infidelity, while photon loss in the retained cable modes remains smaller but non-negligible. These results identify DTC-mediated tunable nonlocal $ZZ$ coupling as a promising gate primitive for modular superconducting processors based on fixed-frequency qubits.

\section{Results}\label{sec2}

\subsection{Distributed fixed-frequency architecture with a gate-native interconnect}

The proposed architecture uses tunable couplers, rather than tunable computational qubits, to generate a remote conditional interaction between spatially separated fixed-frequency transmons. This design choice is central to the scheme. It confines flux control to the coupler sector while allowing the qubit frequencies to remain fixed during the gate, thereby separating the gate-control channel from the coherence-protected computational modes. In this sense, the coaxial link is used not as a passive state-transfer channel \cite{RN291}, but as a gate-native element that participates directly in the formation of a switchable nonlocal $ZZ$ interaction.

\begin{figure*}
\centering
\includegraphics[width=0.8\linewidth]{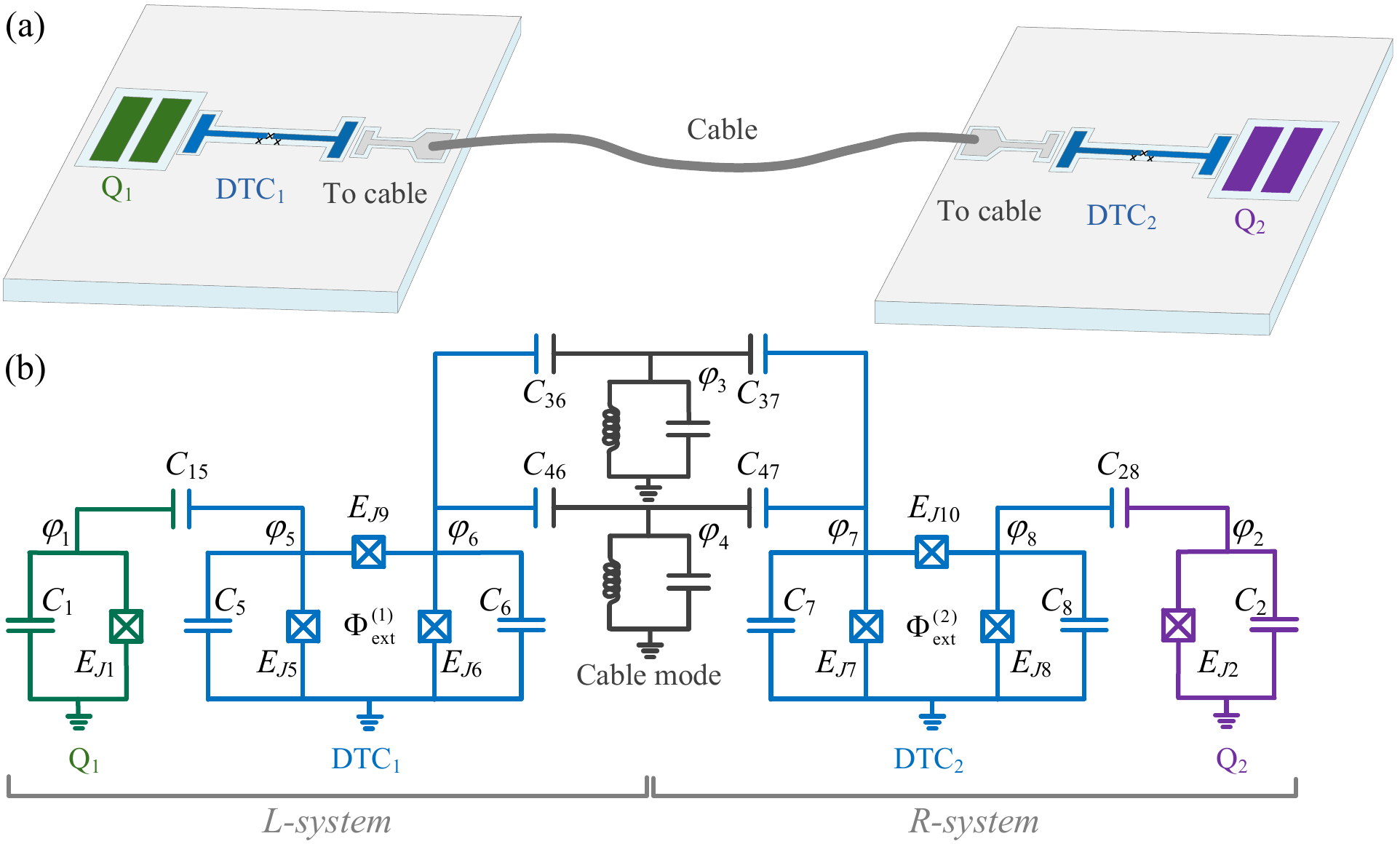}
\caption{\label{Fig1}Schematic of the distributed fixed-frequency architecture with a DTC-mediated gate-native interconnect. (a) Physical layout in which two transmon qubits are housed in separate packages and connected by a 25-cm coaxial cable. On-chip double-transmon couplers (DTCs) and interface structures, labelled ``To cable'', mediate the connection between each qubit module and the cable. (b) Equivalent circuit model. The fixed-frequency transmon qubits $\mathrm{Q}_1$ (green) and $\mathrm{Q}_2$ (purple) are capacitively coupled to a multimode coaxial cable through DTC$_1$ and DTC$_2$ (blue), respectively. Here, $E_{Ji}$ and $C_i$ denote the Josephson energy and capacitance of mode $i$, $E_{J9}$ and $E_{J10}$ denote the Josephson energies of the coupling junctions in the two DTC loops, and $\Phi_{\mathrm{ext}}^{(1)}$ and $\Phi_{\mathrm{ext}}^{(2)}$ are the external fluxes threading the corresponding DTC loops. The variables $\varphi_i$ denote node phases. In the numerical model used below, the 25-cm cable is represented by the two harmonic modes closest to the qubit frequencies. The grey brackets indicate the left and right local subsystems used to analyse the DTC-controlled qubit--cable interaction before constructing the full nonlocal coupling.}
\end{figure*}

As illustrated in Fig.~\ref{Fig1}, two fixed-frequency transmon qubits, $\mathrm{Q}_1$ and $\mathrm{Q}_2$, are housed in separate modules and are connected through a shared 25-cm coaxial cable. Each qubit couples capacitively to a local double-transmon coupler, denoted by DTC$_1$ and DTC$_2$, and the two DTCs couple to the relevant modes of the cable. This qubit--DTC--cable--DTC--qubit layout differs from direct qubit--cable coupling schemes in that the tunability is inserted between each qubit and the cable. The remote interaction is therefore controlled by the external fluxes applied to the DTCs, rather than by tuning the qubits themselves.

Each DTC consists of two transmon modes connected by a flux-threaded Josephson loop. The external flux $\Phi_{\mathrm{ext}}^{(i)}$ changes the spectrum and nonlinear response of DTC$_i$, which in turn modulates the effective cross-Kerr interaction between the adjacent fixed-frequency qubit and the cable modes. This local flux-controlled interaction is the elementary switching mechanism of the architecture. When both local switches are biased near their operation points, the two qubit--cable interactions combine to produce an effective nonlocal $ZZ$ interaction between $\mathrm{Q}_1$ and $\mathrm{Q}_2$. When the DTCs are biased near their idle points, the same interaction channel is strongly suppressed.

The macroscopic interconnect is modelled as a multimode $\lambda/2$ transmission-line resonator. For a 25-cm coaxial cable, the free spectral range is approximately $440~\mathrm{MHz}$. We retain the two cable modes closest to the qubit frequencies, labelled $m=10$ and $m=11$, and verify the validity of this truncation by comparing with a larger cable-mode basis (see Methods). This reduced model captures the dominant near-resonant qubit--cable interactions while keeping the full circuit simulation computationally tractable.

The total Hamiltonian used for the circuit-level simulations is written as
\begin{equation}
\hat{H}=
\hat{H}_{\mathrm{q}}+
\hat{H}_{\mathrm{DTC}}+
\hat{H}_{\mathrm{cable}}+
\hat{H}_{\mathrm{int}},
\end{equation}
with
\begin{equation}
\begin{aligned}
\hat{H}_{\mathrm{q}} &=
\sum_{i=1}^{2}
\left(4E_{Ci}\hat{n}_{i}^{2}-E_{Ji}\cos\hat{\varphi}_{i}\right), \\
\hat{H}_{\mathrm{DTC}} &=
\sum_{i=5}^{8}
\left(4E_{Ci}\hat{n}_{i}^{2}-E_{Ji}\cos\hat{\varphi}_{i}\right) \\
&\quad
-E_{J9}\cos\left(\hat{\varphi}_5-\hat{\varphi}_6+2\pi\Phi_{\mathrm{ext}}^{(1)}\right) \\
&\quad
-E_{J10}\cos\left(\hat{\varphi}_7-\hat{\varphi}_8+2\pi\Phi_{\mathrm{ext}}^{(2)}\right), \\
\hat{H}_{\mathrm{cable}} &=
\sum_{i=3}^{4}
\left(4E_{Ci}\hat{n}_{i}^{2}+\frac{1}{2}E_{Li}\hat{\varphi}_{i}^{2}\right), \\
\hat{H}_{\mathrm{int}} &=
\sum_{(l,k)}J_{lk}\hat{n}_{l}\hat{n}_{k}.
\end{aligned}
\end{equation}
Here, $\hat{n}_i$ and $\hat{\varphi}_i$ are conjugate Cooper-pair number and phase operators satisfying $[\hat{\varphi}_i,\hat{n}_j]=i\delta_{ij}$. The first three terms describe the fixed-frequency qubits, the two DTCs and the retained cable modes. The interaction Hamiltonian contains the capacitive couplings between connected circuit nodes, including the qubit--DTC and DTC--cable couplings. The parameters $E_{Ci}$, $E_{Ji}$ and $E_{Li}$ are the charging, Josephson and cable-mode inductive energies, respectively. The coupling strengths $J_{lk}$ are obtained from the circuit capacitance matrix, and the full parameter set is listed in Table~\ref{tab1}.

The two retained cable modes have different spatial parities along the transmission line, so their coupling amplitudes to the two spatially separated DTCs can differ in sign. This sign structure is included explicitly in the circuit model. In the parameter set used here, the coupling of mode $m=11$ to the right DTC has the opposite sign to its coupling to the left DTC; specifically, we use $J_{46}^{(m=11)}=+25~\mathrm{MHz}$ and $J_{47}^{(m=11)}=-25~\mathrm{MHz}$. This mode-dependent sign is important for constructing the full multimode interaction model and is retained in all simulations below.

\subsection{Spectral mechanism of DTC-mediated switchable nonlocal $ZZ$ interaction}

\begin{figure*}
\centering
\includegraphics[width=1\linewidth]{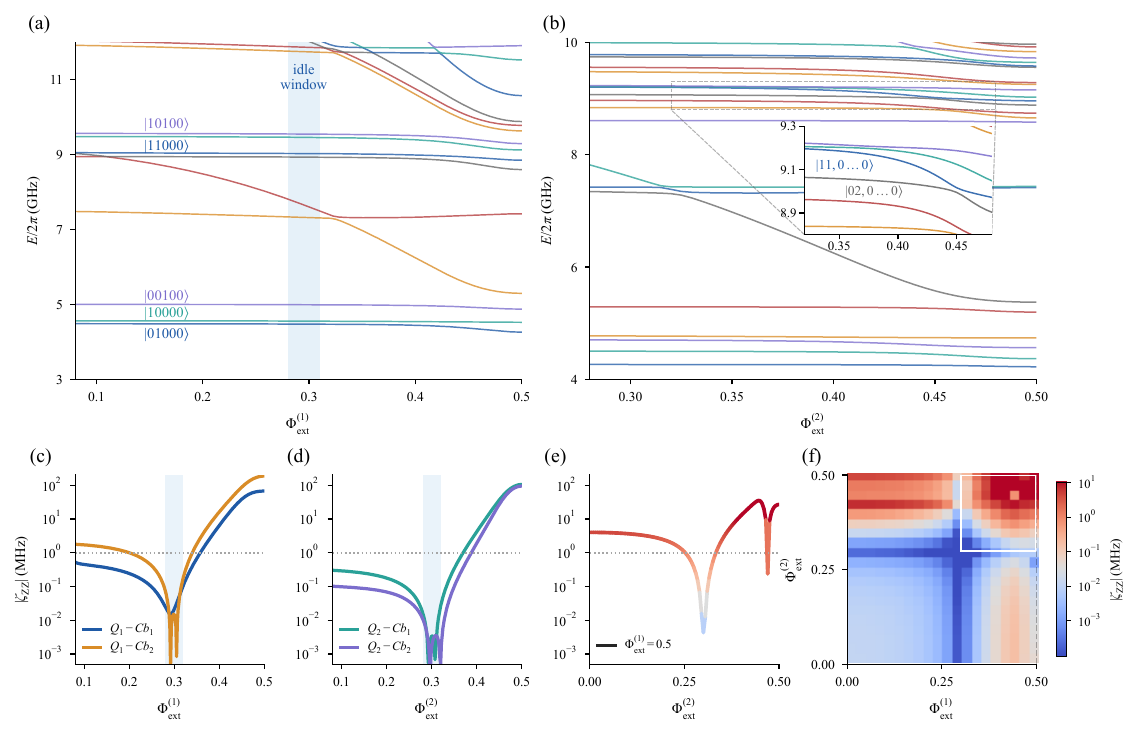}
\caption{\label{Fig2}Spectral mechanism of DTC-mediated switchable nonlocal $ZZ$ interaction. 
(a) Flux-dependent energy spectrum of a local subsystem consisting of qubit $Q_1$, DTC$_1$, and the two retained cable modes $m=10$ and $m=11$. The eigenstates are labelled by their dominant bare-state components, $|Q_1,Cb_1,Cb_2,Cp_{1A},Cp_{1B}\rangle$, where $Cb_1$ and $Cb_2$ denote the two cable modes and $Cp_{1A}$ and $Cp_{1B}$ denote the two internal modes of DTC$_1$. 
(b) Energy spectrum of the full eight-mode system as a function of the flux applied to DTC$_2$, with the flux of DTC$_1$ fixed near its operation point $\Phi_\mathrm{ext}^{1}=0.5$. The inset highlights the avoided-crossing region between the computational two-excitation state $|11,0\dots0\rangle$ and the $|02,0\dots0\rangle$-dominated dressed branch, which provides the main spectral pathway for conditional-phase accumulation; nearby non-computational channels are monitored as leakage pathways.
(c) Effective cross-Kerr couplings between $Q_1$ and the two retained cable modes as functions of $\Phi_{\mathrm{ext}}^{(1)}$. The two curves correspond to the $Q_1$--$Cb_1$ and $Q_1$--$Cb_2$ couplings, showing that DTC$_1$ can suppress or activate the local qubit-cable interaction. 
(d) Effective cross-Kerr couplings between $Q_2$ and the two retained cable modes as functions of $\Phi_{\mathrm{ext}}^{(2)}$. The analogous flux dependence demonstrates that DTC$_2$ provides a second local qubit-cable switch at the opposite end of the cable. 
(e) One-dimensional cut of the effective inter-qubit $ZZ$ coupling with one DTC held near its operation point. This cut shows that the nonlocal interaction can be tuned by the flux applied to the other DTC. 
(f) Two-dimensional landscape of the effective nonlocal $ZZ$ coupling under joint flux control of $\Phi_{\mathrm{ext}}^{(1)}$ and $\Phi_{\mathrm{ext}}^{(2)}$. Blue and red regions correspond to low- and high-coupling regimes, respectively, illustrating the tunability of the DTC-mediated remote interaction. The white outline marks the usable operating window. The rightmost black dashed line denotes an off-configuration where tuning DTC$_1$ suppresses the nonlocal coupling even when DTC$_2$ remains in the high-coupling regime, consistent with the single-DTC switch-off behaviour in (e).}
\end{figure*}

To identify how the DTCs activate a remote conditional interaction, we separate the spectral analysis into local and global levels. The local analysis isolates the flux-controlled qubit-cable interaction generated by a single DTC, whereas the global analysis shows how the two local interactions combine into an effective nonlocal $ZZ$ coupling between the remote qubits. The local qubit--cable cross-Kerr terms identify the flux regions in which each DTC activates its end of the interconnect, whereas the nonlocal $ZZ$ interaction is extracted directly from the dressed computational energies of the full multimode circuit.

We first consider the left local subsystem, denoted by the $L$ system, which contains qubit $Q_1$, DTC$_1$, and the two retained cable modes $m=10$ and $m=11$. This subsystem is used to determine how the flux applied to DTC$_1$ controls the local qubit-cable cross-Kerr interaction. Consistent with the circuit Hamiltonian introduced above, the local Hamiltonian can be written as
\begin{equation}
\hat{H}_L
=
\hat{H}_{q_1}
+
\hat{H}_{\mathrm{DTC}_1}
+
\sum_{\alpha=3}^{4}
\hat{H}_{\mathrm{cable},\alpha}
+
\hat{H}_{\mathrm{int},L},
\end{equation}
with
\begin{equation}
\begin{aligned}
\hat{H}_{q_1}
&=
4E_{C1}\hat{n}_{1}^{2}
-
E_{J1}\cos\hat{\varphi}_{1}, \\
\hat{H}_{\mathrm{DTC}_1}
&=
\sum_{i=5}^{6}
\left(
4E_{Ci}\hat{n}_{i}^{2}
-
E_{Ji}\cos\hat{\varphi}_{i}
\right) \\
&\quad -
E_{J9}\cos\left(
\hat{\varphi}_{5}
-
\hat{\varphi}_{6}
+
2\pi\Phi_{\mathrm{ext}}^{(1)}
\right), \\
\hat{H}_{\mathrm{cable},\alpha}
&=
4E_{C\alpha}\hat{n}_{\alpha}^{2}
+
\frac{1}{2}E_{L\alpha}\hat{\varphi}_{\alpha}^{2},
\quad (\alpha=3,4), \\
\hat{H}_{\mathrm{int},L}
&=
J_{15}\hat{n}_{1}\hat{n}_{5}
+
J_{56}\hat{n}_{5}\hat{n}_{6} \\
&\quad +
\sum_{\alpha=3}^{4}
J_{\alpha6}\hat{n}_{\alpha}\hat{n}_{6}.
\end{aligned}
\end{equation}
Here, the two cable indices $\alpha=3,4$ correspond to the retained modes $m=10$ and $m=11$, labelled as $Cb_1$ and $Cb_2$, respectively. The interaction term contains the local qubit-DTC coupling, the internal DTC coupling, and the DTC-cable couplings that mediate the flux-dependent qubit-cable interaction.

The eigenenergies and eigenstates of the $L$ system are denoted by $E_{Q_1,Cb_1,Cb_2,Cp_{1A},Cp_{1B}}$and $|Q_1,Cb_1,Cb_2,Cp_{1A},Cp_{1B}\rangle$,
where the indices specify the dominant bare-state occupation numbers of the qubit, the two cable modes, and the two internal DTC modes. For a selected cable mode $Cb_\alpha$, the effective qubit-cable cross-Kerr coupling is extracted from the dressed energy shifts as
\begin{equation}
\begin{aligned}
\zeta_{Q_i,Cb_\alpha}
&=
E_{1,1_\alpha,0_{\bar{\alpha}},0,0}
-
E_{1,0_\alpha,0_{\bar{\alpha}},0,0} \\
&\quad -
E_{0,1_\alpha,0_{\bar{\alpha}},0,0}
+
E_{0,0_\alpha,0_{\bar{\alpha}},0,0}.
\end{aligned}
\end{equation}
Here, $i=1,2$ labels the qubit, $1_\alpha$ denotes one excitation in the selected cable mode, $0_{\bar{\alpha}}$ denotes zero excitation in the other retained cable mode, and the two DTC modes remain in their ground states. This energy combination measures the conditional shift associated with jointly exciting a qubit and a cable mode.

Because the flux-dependent spectrum contains several avoided crossings, the state labels must be tracked consistently as the external flux is varied. We assign each dressed eigenstate by its maximum overlap with the corresponding bare basis state,
\begin{equation}
|\psi_i^{\mathrm{dressed}}\rangle
=
\underset{|\psi_j\rangle}{\mathrm{argmax}}
\,
|\langle\psi_i^{\mathrm{bare}}|\psi_j\rangle|^2 .
\end{equation}
This overlap-based procedure allows the qubit-like, cable-like, and coupler-like states to be followed across the flux range and provides a consistent basis for extracting both local and nonlocal $ZZ$ couplings.

The local spectrum in Fig.~\ref{Fig2}(a) shows how DTC$_1$ reshapes the dressed-state structure of the $L$ system. From these dressed energies, we extract the local $Q_1$-cable cross-Kerr couplings shown in Fig.~\ref{Fig2}(c). Both the $Q_1$--$Cb_1$ and $Q_1$--$Cb_2$ couplings are strongly controlled by $\Phi_{\mathrm{ext}}^{(1)}$: near $\Phi_{\mathrm{ext}}^{(1)}\approx0.3$, the couplings are suppressed, defining a local idle point, whereas near $\Phi_{\mathrm{ext}}^{(1)}\approx0.5$, the coupling strengths increase substantially, defining the operation region for DTC$_1$.

The same analysis applies to the right local subsystem, or $R$ system, which contains $Q_2$, DTC$_2$, and the retained cable modes. Its Hamiltonian has the analogous form
\begin{equation}
\hat{H}_R
=
\hat{H}_{q_2}
+
\hat{H}_{\mathrm{DTC}_2}
+
\sum_{\alpha=3}^{4}
\hat{H}_{\mathrm{cable},\alpha}
+
\hat{H}_{\mathrm{int},R}.
\end{equation}
As shown in Fig.~\ref{Fig2}(d), DTC$_2$ similarly controls the $Q_2$--$Cb_1$ and $Q_2$--$Cb_2$ cross-Kerr couplings. The right-side couplings are suppressed near $\Phi_{\mathrm{ext}}^{(2)}\approx0.3$ and become large near $\Phi_{\mathrm{ext}}^{(2)}\approx0.5$. Thus, the two DTCs act as independently tunable local switches at the two ends of the cable.

Having identified the two local switching elements, we next construct the full eight-mode Hamiltonian containing two qubits, four DTC modes, and the two retained cable modes. The global eigenenergies and eigenstates are denoted by $E_{\vec{n}}$ and $|\psi_{\vec{n}}\rangle$, where
$\vec{n}\equiv\{n_{Q_1},n_{Q_2},n_{Cb_1},n_{Cb_2},n_{Cp_{1A}},n_{Cp_{1B}},n_{Cp_{2A}},n_{Cp_{2B}}\}$
specifies the excitation number in each constituent mode. The full-system spectrum in Fig.~\ref{Fig2}(b) shows the avoided-crossing structure relevant to the nonlocal conditional interaction. 

The effective nonlocal $ZZ$ coupling is extracted from the dressed computational energies as
\begin{equation}
\zeta_{Q_1Q_2}=E_{11}-E_{10}-E_{01}+E_{00},
\end{equation}
where the cable and DTC modes are assigned to their corresponding dressed ground-state branches. Fig.~\ref{Fig2}(e) shows a one-dimensional cut of $\zeta_{Q_1Q_2}$ with one DTC held near its operation point while the other flux is varied. The inter-qubit $ZZ$ coupling remains strongly tunable in this configuration, showing that the remote interaction can be controlled by either DTC once the opposite side is activated.

Fig.~\ref{Fig2}(f) presents the full two-dimensional landscape of $\zeta_{Q_1Q_2}$ under joint modulation of $\Phi_{\mathrm{ext}}^{(1)}$ and $\Phi_{\mathrm{ext}}^{(2)}$. When both DTCs are biased near their idle points, $\Phi_{\mathrm{ext}}^{(1)}\approx\Phi_{\mathrm{ext}}^{(2)}\approx0.3$, the inter-qubit $ZZ$ coupling is suppressed to the $10^{-5}$~MHz scale, consistent with weak residual coupling during remote-qubit idling. When both DTCs are moved toward the operation region near $\Phi_{\mathrm{ext}}^{(1)}\approx\Phi_{\mathrm{ext}}^{(2)}\approx0.5$, the interaction increases to the MHz-to-10-MHz scale. This flux-controlled contrast between the idle and operation regimes establishes the DTC-mediated nonlocal $ZZ$ interaction used for the remote-CZ dynamics analysed below.

\subsection{Remote controlled-Z gate dynamics}

The remote CZ gate is implemented by synchronously modulating the external fluxes of the two DTCs. Based on the coupling analysis above, the effective inter-qubit $ZZ$ interaction is suppressed to the $10^{-5}$~MHz scale when both couplers are biased near the idle point, $\Phi_{\mathrm{idle}}\approx0.3$. We therefore define the computational basis states $|ij,0\dots0\rangle$ $(i,j\in\{0,1\})$ as the dressed eigenstates of the full circuit at this idle bias. To execute the gate, the fluxes are pulsed towards the operation region near $\Phi_{\mathrm{work}}\approx0.5$, where the nonlocal $ZZ$ interaction is enhanced. In this region, the dressed branch connected to $|\widetilde{11}\rangle$ approaches a branch with dominant $|\widetilde{02}\rangle$ character, producing an avoided crossing that shifts the $|\widetilde{11}\rangle$ energy relative to the other computational branches. The resulting conditional dynamical phase realizes a controlled-phase operation when the population returns to the computational subspace at the end of the pulse. As shown in Fig.~\ref{Fig2}(f), the white outline marks the usable operating window. In the simulations below, we use the full circuit Hamiltonian rather than a reduced effective model, thereby retaining the relevant multimode dressing and leakage channels. All device parameters are listed in Table~\ref{tab1} of Methods, and the time evolution is solved using the QuTiP solver~\cite{qutip}.

\begin{figure*}
\centering
\includegraphics[width=1\linewidth]{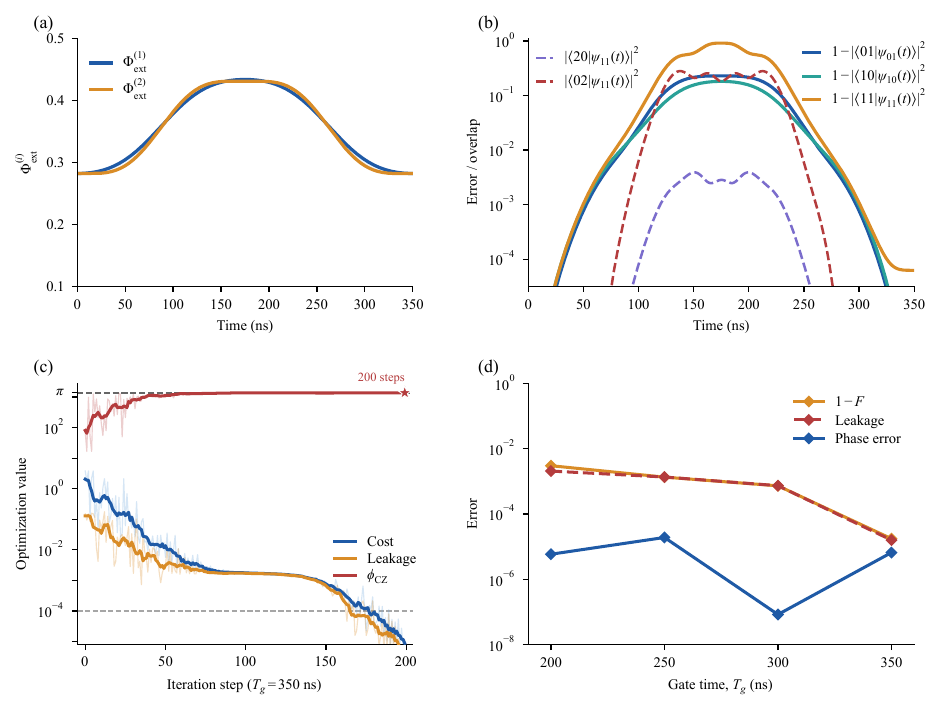}
\caption{\label{Fig3}Remote-CZ pulse and coherent gate dynamics. The optimized synchronous DTC flux pulse activates the nonlocal interaction and converts it into a controlled-phase operation between remote fixed-frequency qubits. (a) Flux waveforms applied simultaneously to the two DTCs. (b) Population dynamics during the optimized pulse, showing the computational-state populations and the transient occupations of the dominant non-computational channels, including the $|02\rangle$- and $|20\rangle$-derived dressed branches. (c) Optimization trajectory of the pulse parameters used to accumulate the target conditional phase while suppressing leakage and unwanted population transfer.
(d)  Projected average gate fidelity, leakage error and conditional-phase error as functions of gate duration under coherent evolution.}
\end{figure*}

To implement the gate operation, we modulate the coupler flux using a smooth waveform constructed from a truncated Fourier series,
\begin{equation}
\Phi(t)
=
\Phi_i
+
\Phi_f
-
\frac{\Phi_f}{2}
\sum_{k=1}^{n}
\lambda_k
\left[
1-\cos\left(
2k\pi\frac{t-T/2}{T}
\right)
\right],
\end{equation}
where $\Phi_i$ is the idle flux bias, $\Phi_f$ denotes the modulation amplitude from the idle point towards the interaction region, $T$ is the gate duration, $\{\lambda_k\}$ are the waveform coefficients to be optimized, and $n$ is the truncation order. The smooth Fourier form suppresses abrupt spectral excursions while allowing the pulse to approach the interaction region in which the conditional phase is accumulated.

To optimize the remote CZ gate, we minimize leakage out of the computational subspace while enforcing the target conditional phase. The leakage error is quantified as $L=1-\langle\psi_f|\hat{P}_\mathcal{C}|\psi_f\rangle$
where $|\psi_f\rangle$ is the final state evolved from an initial computational state $|\psi_0\rangle=\sum_{i,j=0,1}c_{ij}|ij\rangle$, and
$\hat{P}_\mathcal{C}=\sum_{i,j=0,1}|ij\rangle\langle ij|$ is the projector onto the computational subspace $\mathcal{C}$. The conditional-phase error is defined as$\delta\phi=\phi_{11}-\phi_{01}-\phi_{10}+\phi_{00}-\pi$ where $\phi_{ij}$ denotes the dynamical phase accumulated by the computational state $|ij\rangle$. We then optimize the pulse parameters to reduce both leakage and phase error. The detailed optimization procedure is presented in the Methods, with the optimization trajectory illustrated in Fig.~\ref{Fig3}(c). To assess the resulting coherent gate performance, we compute the average gate fidelity using~\cite{Fidelity}
\begin{equation}
\overline{F}
=
\frac{
\left|\mathrm{Tr}(\hat{U}_{\mathrm{id}}^\dagger\hat{U})\right|^2
+
\mathrm{Tr}(\hat{U}^\dagger\hat{U})
}
{d(d+1)} ,
\end{equation}
where $\hat{U}_{\mathrm{id}}=\mathrm{diag}(1,1,1,-1)$ is the ideal CZ propagator, and $\hat{U}$ is the actual evolution operator projected onto the two-qubit computational subspace spanned by $\{|00\rangle,|01\rangle,|10\rangle,|11\rangle\}$, with $d=4$. Because leakage makes the projected propagator non-unitary, the term $\mathrm{Tr}(\hat{U}^\dagger\hat{U})$ accounts for the population retained in the computational subspace.

The optimized flux waveform $\Phi(t)$ is shown in Fig.~\ref{Fig3}(a), and the corresponding coherent dynamics are shown in Fig.~\ref{Fig3}(b). To quantify unwanted population transfer within the single-excitation manifold, we initialize the system in $|\widetilde{01}\rangle$ and $|\widetilde{10}\rangle$ and define the swap errors as $\varepsilon_\mathrm{swap}^{01}=1-P_{01}$ and $\varepsilon_\mathrm{swap}^{10}=1-P_{10}$, respectively, where $P_{ij}$ denotes the population in the computational state $|\widetilde{ij}\rangle$. Both swap errors remain below $10^{-4}$, indicating that the pulse induces negligible population exchange within the single-excitation computational manifold. We further initialize the system in $|\widetilde{11}\rangle$ and define the final leakage error as $\varepsilon_{\mathrm{leak}}=1-P_{11}$. This leakage error also remains below $10^{-4}$ at the end of the pulse.

The transient dynamics from the $|\widetilde{11}\rangle$ initial state reveal how the conditional phase is accumulated. During the pulse, population is temporarily transferred from the $|\widetilde{11}\rangle$-derived branch to the dominant non-computational channels, including $|\widetilde{02}\rangle$ and $|\widetilde{20}\rangle$. The larger contribution is associated with the $|\widetilde{02}\rangle$ channel, consistent with the avoided-crossing structure of the full-system spectrum. This transient hybridization shifts the $|\widetilde{11}\rangle$-derived energy and accumulates the required conditional phase, while the shaped pulse returns the population to the computational subspace by the end of the evolution. For a gate duration of $T=350$~ns, the conditional phase converges to $\phi_{\mathrm{CZ}}\approx\pi$, yielding an average coherent gate fidelity of $F=99.99\%$.

Fig.~\ref{Fig3}(c) shows the optimization process for the waveform coefficients, and Fig.~\ref{Fig3}(d) summarizes the dependence of the gate fidelity, leakage error and conditional-phase error on gate duration. These coherent-dynamics simulations show that the DTC-mediated nonlocal $ZZ$ interaction can be converted into a remote CZ gate with low coherent leakage under an optimized flux pulse. The effects of decoherence and coherence-time constraints are analysed separately in the following section.

\subsection{Decoherence-limited fidelity}

While idealized unitary dynamics capture the Hamiltonian mechanism of the remote gate, realistic superconducting circuits are coupled to dissipative environments during the driven evolution. We therefore assess the decoherence-limited performance of the optimized pulse within a Markovian open-system model. The model includes finite relaxation and dephasing times for the endpoint qubits and DTC modes, together with photon loss in the retained coaxial-cable modes.

\begin{figure*}
\centering
\includegraphics[width=1\linewidth]{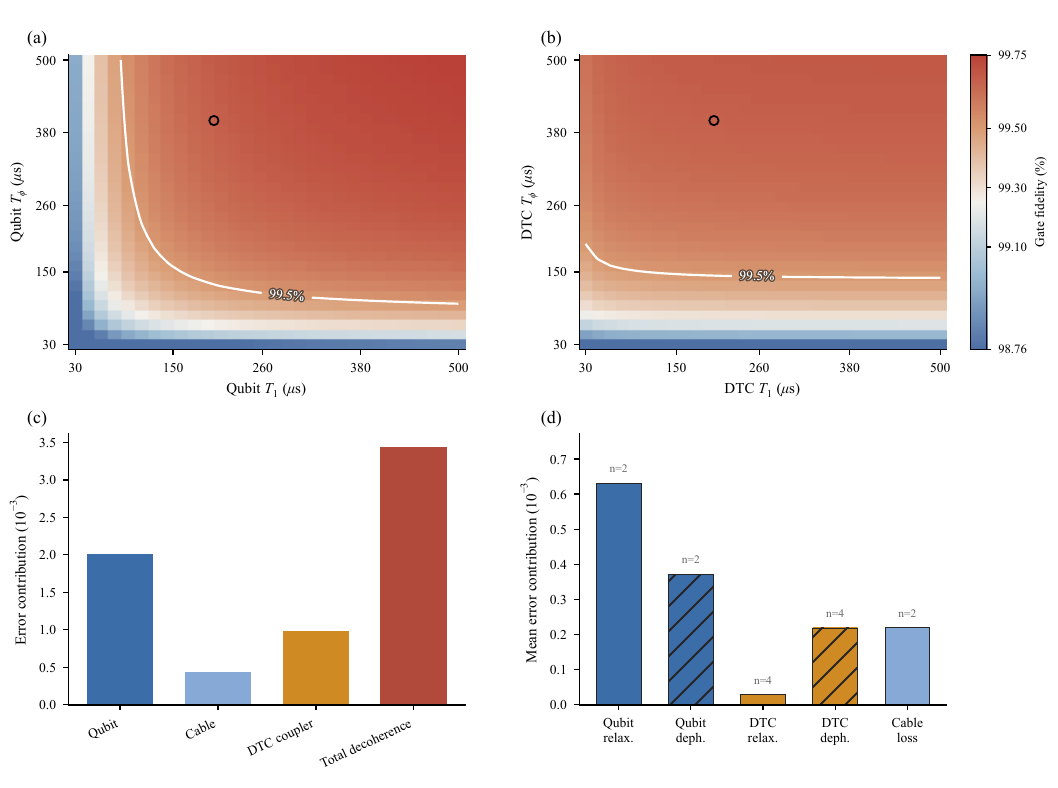}
\caption{
\label{Fig4}
Decoherence-limited remote-CZ fidelity under representative Markovian noise.
(a) Average gate fidelity versus qubit coherence time, evaluated with the optimized coherent pulse while fixing the DTC coherence parameters at $T_{1,c}=200~\mu{\rm s}$ and $T_{\phi,c}=400~\mu{\rm s}$ and the retained cable modes at $Q_{\rm cable}=10^6$. The white contour marks the $99.5\%$ reference fidelity threshold.
(b) Average gate fidelity versus DTC coherence time, with the qubit parameters fixed at $T_{1,q}=200~\mu{\rm s}$ and $T_{\phi,q}=400~\mu{\rm s}$ and the cable quality factors fixed at $Q_{\rm cable}=10^6$, isolating the sensitivity to coupler decoherence.
(c) Component-resolved fidelity loss for the representative parameter set $T_{1,q}=200~\mu{\rm s}$, $T_{\phi,q}=400~\mu{\rm s}$, $T_{1,c}=200~\mu{\rm s}$, $T_{\phi,c}=400~\mu{\rm s}$ and $Q_{\rm cable}=10^6$, obtained from Lindblad master-equation simulations using the optimized coherent pulse. Under these conditions, the decoherence-limited fidelity reaches approximately $99.65\%$, corresponding to a total infidelity of about $0.34\%$. The summed error budget assigns $58.3\%$ of the infidelity to endpoint-qubit relaxation and dephasing, $28.7\%$ to DTC relaxation and dephasing, and $12.8\%$ to photon loss in the retained cable modes, indicating that cable photon loss is a smaller but non-negligible contribution in this Markovian model.
(d) Mode-averaged error contribution for each physical channel class at the representative parameter set used in (c).
}
\end{figure*}

The decoherence effects in the open quantum system are formally captured by the Lindblad dissipator superoperator, defined as \cite{Lind}
\begin{equation}
\mathcal{D}[\hat{L}_k]\hat{\rho} = \hat{L}_k \hat{\rho} \hat{L}_k^\dagger - \frac{1}{2}\left( \hat{L}_k^\dagger \hat{L}_k \hat{\rho} + \hat{\rho} \hat{L}_k^\dagger \hat{L}_k \right),
\end{equation}
which acts on the system density matrix $\hat{\rho}$. Taking into account $N_L$ distinct dissipation channels, the overall time evolution of the system is governed by the Lindblad master equation \cite{Universal-fidelity}:
\begin{equation}
\dot{\hat{\rho}}(t) = -\frac{i}{\hbar}\left[\hat{H}(t),\hat{\rho}(t)\right] + \sum_{k=1}^{N_L}\Gamma_k\mathcal{D}\bigl[\hat{L}_k\bigr]\hat{\rho}(t),
\label{eq:lindblad_master}
\end{equation}
where $\hat{L}_k$ denotes the jump operator characterizing the $k$-th dissipation process, and $\Gamma_k$ is the corresponding decay rate. For each transmon-like mode $j$, energy relaxation and pure dephasing are included with rates
\begin{equation}
\Gamma_{1,j}=\frac{1}{T_{1,j}},
\qquad
\Gamma_{\phi,j}=\frac{1}{T_{\phi,j}}.
\end{equation}
Photon loss in cable mode $m$ is included with rate
\begin{equation}
\kappa_m=\frac{\omega_m}{Q_m}.
\end{equation}

By expanding the master equation perturbatively in the dissipation rates \cite{Perturbation}, the first-order contribution from each dissipative channel can be written as
\begin{equation}
\label{fidelity}
\overline{F} = 1 + \sum_{k=1}^{N_L} \Gamma_k \int_{0}^{\tau} dt \, \delta F(t, \hat{L}_k) + \mathcal{O}(\tau^2 \Gamma_k^2),
\end{equation}
where $\tau$ is the gate duration and $\delta F_k(t)$ denotes the instantaneous fidelity correction associated with the $k$-th dissipation channel. For a jump operator $\hat{L}_k$, this correction is given by \cite{CABRERA200725,Abad2025impactofdecoherence}
\begin{equation}
\begin{split}
\delta F_k(t) = & \frac{1}{d(d+1)} \left| \operatorname{Tr}_{\rm cmp} \left[ \hat{L}_k(t) \right] \right|^2 \\
& - \frac{1}{d+1} \operatorname{Tr}_{\rm cmp} \left[ \hat{L}_k^\dagger(t) \hat{L}_k(t) \right],
\end{split}
\label{eq:delta_fidelity}
\end{equation}
with $\hat{L}_k(t)=\hat{U}^\dagger(t)\hat{L}_k\hat{U}(t).$ Here, $\hat{U}(t)$ is the coherent evolution operator generated by the time-dependent Hamiltonian $\hat{H}(t)$ in the absence of dissipation, $d=4$ is the dimension of the two-qubit computational subspace, and $\operatorname{Tr}_{\rm cmp}$ denotes a trace restricted to that subspace.

To make the perturbative error estimate physically interpretable, we assigned the jump operators to the hardware modes that participate in the gate. For the two computational qubits and the four DTC modes, energy relaxation is described by the lowering operators $\hat{L}_{1,j}=\hat{b}_j$, and pure dephasing is described by the number operators $\hat{L}_{\phi,j}=\hat{b}_j^\dagger\hat{b}_j$, where $\hat{b}_j$ denotes the annihilation operator of the corresponding transmon-like mode in the truncated basis. Photon loss in the coaxial link is described by $\hat{L}_{m}=\hat{a}_m$ for the retained cable modes $m=10$ and $m=11$, whose frequencies are $\omega_{10}/2\pi=4.5~{\rm GHz}$ and $\omega_{11}/2\pi=4.95~{\rm GHz}$, respectively. The optimized remote-CZ pulse has a duration of $\tau=350~{\rm ns}$. With these assignments, the first-order correction in Eq.~\eqref{fidelity} can be decomposed into endpoint-qubit relaxation and dephasing, DTC-mode relaxation and dephasing, and retained-cable-mode photon-loss contributions.

We first examined how the decoherence-limited fidelity depends on the endpoint-qubit coherence times. In Fig.~\ref{Fig4}(a), $T_{1,q}$ and $T_{\phi,q}$ are varied over $[30,500]~\mu{\rm s}$, while the DTC and cable parameters are fixed at $[T_{1,c}=200~\mu{\rm s}, T_{\phi,c}=400~\mu{\rm s}, Q_{\rm cable}=10^6]$. The average CZ fidelity increases monotonically with both qubit relaxation and dephasing times and exceeds the $99.5\%$ reference contour in the region where both $T_{1,q}$ and $T_{\phi,q}$ are above approximately $160~\mu{\rm s}$. This dependence is expected because the computational population remains predominantly in the two-qubit subspace during the pulse, so endpoint-qubit relaxation and dephasing directly reduce the projected gate fidelity.

We then repeat the scan for the coherence times of the DTC modes while keeping the endpoint-qubit coherence fixed at $[T_{1,q}=200~\mu{\rm s}, T_{\phi,q}=400~\mu{\rm s}]$. As shown in Fig.~\ref{Fig4}(b), the fidelity is less sensitive to the DTC coherence than to the qubit coherence over the parameter range considered. For $T_{1,c}$ and $T_{\phi,c}$ above $150~\mu{\rm s}$, the calculated fidelity remains above $99.5\%$. This weaker dependence is consistent with the role of the DTC as a tunable mediator: the coupler modes activate the nonlocal conditional interaction, but their population remains transient during the optimized gate. Thus, DTC decoherence contributes to the calculated error budget, but it is not the dominant contribution for the operating point used here.

The component-resolved error budget provides a quantitative decomposition of the decoherence-limited infidelity. At the representative operating point marked by the black circle in \ref{Fig4}(a,b)), with $T_{1,q}=200~\mu{\rm s}$, $T_{\phi,q}=400~\mu{\rm s}$, $T_{1,c}=200~\mu{\rm s}$, $T_{\phi,c}=400~\mu{\rm s}$ and $Q_{\rm cable}=10^6$, the decoherence-limited average CZ fidelity is $\overline{F}=99.65\%$, corresponding to a total infidelity of $1-\overline{F}=0.34\%$ (Fig.~\ref{Fig4}(c)). In this summed component-level budget, endpoint-qubit relaxation and dephasing account for $58.3\%$ of the infidelity, while relaxation and dephasing of the DTC modes account for $28.7\%$. Photon loss in the retained cable modes contributes the remaining $12.8\%$.

To complement this total error budget, we also compute a mode-averaged contribution for each physical channel class by averaging the microscopic relaxation, dephasing or photon-loss terms over the modes belonging to that class (Fig.~\ref{Fig4}(d)). This per-mode comparison shows that the endpoint-qubit channels have the largest average contribution among the hardware classes considered. Because Fig.~\ref{Fig4}(d) reports per-mode averages rather than summed channel weights, we use it only to compare the typical strength of individual microscopic channels; the total component-level error fractions are reported in Fig.~\ref{Fig4}(c).

These results should be interpreted as a decoherence-limited estimate under a Markovian noise model. The calculation includes relaxation, pure dephasing and photon loss during the driven gate, but does not include calibration errors, pulse distortion, non-Markovian cable effects or flux noise beyond the effective dephasing times used above. Within this stated model and parameter regime, the summed error budget in Fig.~\ref{Fig4}(c) suggests that endpoint-qubit decoherence is the largest contribution to the total infidelity, whereas photon loss in the retained cable modes remains a smaller but non-negligible contribution.

\section{Discussion}\label{sec3}

The architectural division of monolithic quantum processors into macroscopically distributed modules is a likely route toward realizing FTQC. However, a critical bottleneck has historically been the performance degradation introduced at chip-to-chip interfaces. The results presented in this work establish a gate-native control strategy for nonlocal entanglement generation across a 25-cm multimode cable. Rather than treating the macroscopic interconnect as a mere quantum-state-transfer channel, our architecture incorporates the multimode transmission line and the two double-transmon couplers into a dynamically tunable interacting system. In this way, the inter-module link becomes part of the engineered gate Hamiltonian rather than a passive communication channel.

The central feature of this architecture is the ability to control a nonlocal $ZZ$ interaction between fixed-frequency qubits. By tuning the two DTCs, the effective interaction can be switched from an idle value below $10^{-5}~{\rm MHz}$ to the megahertz scale at the operating point. This large on/off contrast is important for modular superconducting processors, where remote couplings must be strong enough for practical gate times but sufficiently suppressed during idle periods to avoid coherent crosstalk across modules. Under optimized coherent dynamics, the activated interaction yields a projected coherent remote-CZ fidelity of $99.99\%$.

Beyond the specific gate simulation, a key implication of this approach is that it provides a gate-native route to nonlocal entanglement generation without relying on direct quantum-state transfer or tunable computational qubits. In state-transfer-based architectures, quantum information must be emitted into an interconnect, transmitted, captured by a remote node and then converted into a two-qubit operation. Each step introduces additional control requirements and potential sensitivity to loss or mode mismatch. By contrast, the present scheme directly engineers a conditional-phase interaction between distant qubits, making it well matched to remote CZ-type operations. At the same time, because tunability is confined to the coupler sector, the coherence and frequency-allocation advantages of fixed-frequency qubits can be largely preserved.

The open-system analysis places the coherent gate performance in a more experimentally relevant context. For the representative parameter set considered in Fig.~\ref{Fig4}, with $T_{1,q}=200~\mu{\rm s}$, $T_{\phi,q}=400~\mu{\rm s}$, $T_{1,c}=200~\mu{\rm s}$, $T_{\phi,c}=400~\mu{\rm s}$, and $Q_{\rm cable}=10^6$, the decoherence-limited average CZ fidelity is $99.65\%$. The component-resolved error budget shows that endpoint-qubit relaxation and dephasing account for $58.3\%$ of the total infidelity, while DTC-mode relaxation and dephasing account for $28.7\%$. Cable photon loss contributes $12.8\%$. Within this Markovian noise model and parameter regime, the calculated error budget assigns the largest contribution to endpoint-qubit decoherence, while photon loss in the retained cable modes remains a smaller but non-negligible contribution.

This error hierarchy is important for assessing the architectural value of the scheme. It indicates that, for the simulated operating point and the representative Markovian noise parameters, photon loss in the retained cable modes contributes a minority of the calculated error budget. The cable modes and DTC modes mediate the nonlocal interaction, and their transient participation is consistent with a smaller calculated decoherence contribution than that of the computational qubits for the simulated operating point. This distinction supports the view that, in this architecture, a multimode interconnect can serve as an active gate element rather than only as a communication bus.

Several steps remain before this mechanism can be translated into a complete experimental modular-gate implementation. The present analysis focuses on the intrinsic Hamiltonian mechanism and its decoherence-limited performance under a Markovian noise model. A full device-level assessment will require experimental validation of the flux-control landscape, the transient participation of coupler and cable modes, and the robustness of the gate under integrated multichip operation.

Taken together, these results identify DTC-mediated tunable nonlocal $ZZ$ coupling as a promising candidate building block for distributed superconducting quantum processors. By combining a multimode interconnect architecture with a gate-native flux-control mechanism, this approach provides a gate primitive for remote entangling operations between fixed-frequency qubits. It may therefore offer a useful design route toward modular quantum computing and inter-module operations in future fault-tolerant architectures.

\section{Methods}\label{sec4}
\subsection{Circuit quantization}
As shown in Fig.~\ref{Fig1}(b), the single DTC circuit model consists of two transmons coupled through a Josephson junction. The Lagrangian of DTC is given by
\begin{equation}\mathcal{L}=K-V,\end{equation}
where the kinetic energy $K$ and potential energy $V$ are explicitly defined as:
\begin{equation}\begin{aligned}&K=\left(\frac{1}{2e}\right)^{2}\left[\sum_{i=5,6}\frac{C_{i}}{2}\dot{\varphi}_{i}^{2}+\frac{C_{56}}{2}\left(\dot{\varphi}_{5}-\dot{\varphi}_{6}\right)^{2}\right],\\&V=-\sum_{i=5,6}E_{Ji}\cos\varphi_{i}-E_{J9}\cos\left(\varphi_{5}-\varphi_{6}+2\pi\Phi_\mathrm{ext}^{(1)}\right).\end{aligned}\end{equation}
Here, the generalized node flux $\phi_{i}$ is related to the superconducting phase via $\varphi_i=(2\pi/\Phi_0)\phi_i\mathrm{~,}$ with $\Phi_0=h/2e$ being the magnetic flux quantum. The generalized momenta, which are the canonical conjugates to the node fluxes, correspond to the node charges $q_i\equiv\frac{\partial\mathcal{L}}{\partial\dot{\varphi}_i},\quad i=5,6$. The system Hamiltonian is then obtained via a Legendre transformation ($\hbar = 1$):
\begin{equation}
\begin{aligned}
H &= \sum_{i=5,6} \frac{\partial L}{\partial\dot{\varphi}_{i}} \dot{\varphi}_{i} - \mathcal{L} \\
  &= \left( \frac{1}{2e} \right)^{2} 
     \Bigg[ \sum_{i=5,6} \frac{C_{i}}{2} \dot{\varphi}_{i}^{2} 
          + \frac{C_{56}}{2} \left( \dot{\varphi}_{5} - \dot{\varphi}_{6} \right)^{2} \Bigg] \\
  &\quad - \sum_{i=5,6} E_{Ji} \cos\varphi_{i} 
     - E_{J9} \cos\left( \varphi_{5} - \varphi_{6} + 2\pi\Phi_\mathrm{ext}^{(1)} \right).
\end{aligned}
\end{equation}
It is convenient to rewrite the kinetic term in matrix form as $K=\frac{1}{2}\vec{\dot{\phi}}^T\mathbf{C}\vec{\dot{\phi}},$ where the flux vector is defined as $\vec{\dot{\phi}}=(\dot{\phi}_5,\dot{\phi}_6)^T$ with $\dot{\phi}_i=\frac{1}{2e}\dot{\varphi}_i$. The capacitance matrix $C$ is given by:
\begin{equation}\mathbf{C}=\begin{pmatrix}C_5+C_{56}&-C_{56}\\-C_{56}&C_6+C_{56}\end{pmatrix}.\end{equation}
Using the canonical relation $\vec{q}=\mathbf{C}\vec{\dot{\phi}}$ and assuming the coupling capacitance is small relative to the shunt capacitances (i.e., $C_{56}\ll C_5,C_6$),  the kinetic energy may be written as
\begin{equation}K=\frac{1}{2}\vec{q}^{T}C^{-1}\vec{q},\quad \mathbf{C}^{-1}=\begin{pmatrix}\frac{1}{C_{5}}&\frac{C_{56}}{C_{5}C_{6}}\\\frac{C_{56}}{C_{5}C_{6}}&\frac{1}{C_{6}}\end{pmatrix}.\end{equation}
Substituting this inverse capacitance matrix yields
\begin{equation}K=\frac{1}{2}\left(\frac{4e^2}{C_5}n_5^2+\frac{8e^2C_{56}}{C_5C_6}n_5n_6+\frac{4e^2}{C_6}n_6^2\right),\end{equation}
where we have introduced the number operators $n_i=q_i/2e$. Finally, combining the kinetic and potential terms, the total Hamiltonian for the DTC takes the form:
\begin{equation}\begin{aligned}{H}&=4E_{C5}\hat{n}_{5}^{2}+4E_{C6}\hat{n}_{6}^{2}+J_{56}\hat{n}_{5}\hat{n}_{6}\\&-\sum_{i=5,6}E_{Ji}\cos\hat{\varphi}_{i}-E_{J9}\cos\left(\hat{\varphi}_{5}-\hat{\varphi}_{6}+2\pi\Phi_\mathrm{ext}^{(1)}\right),\end{aligned}\end{equation}
where $E_{Ci}=e^2/2C_i$ represents the charging energy, and $J_{56}=4e^2C_{56}/(C_5C_6)$ denotes the effective coupling strength. This derived Hamiltonian serves as the foundation for our numerical simulations.

Having established the model for the isolated DTC, we now extend our analysis to the $L-system$, where the DTC interacts with two adjacent modes (labeled 1 and 4), as illustrated in Fig.~\ref{Fig1}(b). Applying the same circuit quantization formalism, the Lagrangian for this coupled subsystem is given by:
\begin{equation}
\begin{aligned}
\mathcal{L} = & \left(\frac{1}{2e}\right)^{2} \Biggl[ \sum_{i\in\{1,4,5,6\}} \frac{C_{i}}{2}\dot{\varphi}_{i}^{2} + \frac{C_{15}}{2}(\dot{\varphi}_1-\dot{\varphi}_5)^2 \\
& + \frac{C_{56}}{2}(\dot{\varphi}_5-\dot{\varphi}_6)^2 + \frac{C_{46}}{2}(\dot{\varphi}_4-\dot{\varphi}_6)^2 \Biggr] \\
& - \frac{1}{2}E_{L4}^{(11)}\varphi_{4}^{2} + \sum_{i\in\{1,5,6\}}E_{Ji}\cos\varphi_{i} \\
& + E_{J9}\cos\left(\varphi_{5}-\varphi_{6}+2\pi\Phi_{\mathrm{ext}}^{(1)}\right).
\end{aligned}
\end{equation}

The kinetic energy term can be compactly expressed in matrix notation as $K=\frac{1}{2}\vec{\dot{\phi}}^T\mathbf{C}\vec{\dot{\phi}},$ where the flux vector is defined as $\vec{\dot{\phi}}=(\dot{\phi}_1,\dot{\phi}_4,\dot{\phi}_5,\dot{\phi}_6)^T$. The capacitance matrix $\mathbf{C}$ takes the form:

\begin{equation}
\resizebox{1\columnwidth}{!}{%
$\mathbf{C}=
\begin{pmatrix}
C_1+C_{15} & 0 & -C_{15} & 0 \\
0 & C_4+C_{46} & 0 & -C_{46} \\
-C_{15} & 0 & C_5+C_{15}+C_{56} & -C_{56} \\
0 & -C_{46} & -C_{56} & C_6+C_{46}+C_{56}
\end{pmatrix}$%
}
\label{eq:Cmatrix}
\end{equation}

To transition to the Hamiltonian formalism, we rewrite the kinetic energy in terms of the canonical charge momenta $\vec{q}$ via $K=\frac{1}{2}\vec{q}^T\mathbf{C}^{-1}\vec{q}.$ The inverse capacitance matrix is formally given by $\mathbf{C}^{-1}=\mathrm{adj}(\mathbf{C})/|\mathbf{C}|.$ Assuming the coupling capacitances are small relative to the shunt capacitances, we approximate the inverse matrix to leading order as:
\begin{equation}
\resizebox{\columnwidth}{!}{%
$\renewcommand{\arraystretch}{2.1} 
\mathbf{C}^{-1} \approx 
\begin{pmatrix}
\dfrac{1}{C_1} & \dfrac{C_{15}C_{46}C_{56}}{C_1C_4C_5C_6} & \dfrac{C_{15}}{C_1C_5} & \dfrac{C_{15}C_{56}}{C_1C_5C_6} \\
\dfrac{C_{15}C_{46}C_{56}}{C_1C_4C_5C_6} & \dfrac{1}{C_4} & \dfrac{C_{45}C_{56}}{C_4C_5C_6} & \dfrac{C_{46}}{C_4C_6} \\
\dfrac{C_{15}}{C_1C_5} & \dfrac{C_{45}C_{56}}{C_4C_5C_6} & \dfrac{1}{C_5} & \dfrac{C_{56}}{C_5C_6} \\
\dfrac{C_{15}C_{56}}{C_1C_5C_6} & \dfrac{C_{46}}{C_4C_6} & \dfrac{C_{56}}{C_5C_6} & \dfrac{1}{C_6}
\end{pmatrix}$%
}
\label{eq:Cinverse}
\end{equation}
where $\mathrm{adj}(\mathbf{C})$ denotes the adjugate matrix, and $|\mathbf{C}|$ is the determinant. Consequently, the quantized Hamiltonian for the subsystem is derived as:
\begin{equation}\begin{aligned}\hat{H}&=\sum_{i=1,4,5,6}4E_{Ci}\hat{n}_{i}^{2}+J_{15}\hat{n}_{1}\hat{n}_{5}+J_{56}\hat{n}_{5}\hat{n}_{6}+J_{46}\hat{n}_{4}\hat{n}_{6}\\&-\sum_{i=1,5,6}E_{Ji}\cos\hat{\varphi}_{i}-E_{J9}\cos\left(\hat{\varphi}_{5}-\hat{\varphi}_{6}+2\pi\Phi_\mathrm{ext}^{(1)}\right)\\&+\frac{1}{2}E_{L4}^{m=11}\hat{\varphi}_4^2,\end{aligned}\end{equation}
where $E_{Ci}=e^2/2C_i$ is the charging energy, and $J_{ij}=4e^2C_{ij}/(C_iC_j)$ represents the capacitive coupling strength.

Finally, by generalizing the local interaction terms derived from the subsystem analysis to the full circuit topology, we construct the total system Hamiltonian:
\begin{equation}\begin{aligned}\hat{H}&=\sum_{i=1}^24E_{Ci}\hat{n}_i^2-E_{Ji}\cos\hat{\varphi}_i+\sum_{i=3}^44E_{Ci}\hat{n}_i^2+\frac{1}{2}E_{Li}\hat{\varphi}_i^2\\&+\sum_{i=5}^84E_{Ci}\hat{n}_i^2-E_{Ji}\cos\hat{\varphi}_i+\sum_{l,k}J_{lk}\hat{n}_l\hat{\mathrm{n}}_k\\&-E_{J9}\cos(\hat{\varphi}_5-\hat{\varphi}_6+2\pi\Phi_\mathrm{ext}^{(1)})\\&-E_{J10}\cos(\hat{\varphi}_7-\hat{\varphi}_8+2\pi\Phi_\mathrm{ext}^{(2)}).\end{aligned}\end{equation}

This global Hamiltonian provides the theoretical basis for the numerical simulations presented in the main text. We represent the nonlinear superconducting modes in terms of bosonic operators by introducing the phase and charge operators as $\hat{\varphi}_{i}=\phi_{i,\mathrm{zpf}}(\hat{a}_{i}^\dagger+\hat{a}_{i}),\hat{n}_{i}=in_{i,\mathrm{zpf}}(\hat{a}_{i}^\dagger-\hat{a}_{i}),$ where $\hat{a}_{i}(\hat{a}_{i}^{\dagger})$ denotes the annihilation (creation) operator and $\phi_{i,\mathrm{zpf}}\left(n_{i,\mathrm{zpf}}\right)$ represents the phase (number) zero-point fluctuation, the system Hamiltonian can be approximated by
\begin{equation}
\begin{aligned}
    \hat{H}' &= \sum_{\substack{i=1 \\ i \notin \{3,4\}}}^{8} \left( \omega_{i}\hat{a}_{i}^{\dagger}\hat{a}_{i} - \frac{\alpha_{i}}{2}(\hat{a}_{i}^{\dagger})^2 \hat{a}_{i}^{2} \right) \\& + \sum_{k \in \{3,4\}} \omega_k\hat{c}_{k}^{\dagger}\hat{c}_k + \sum_{\langle i,j\rangle}g_{ij}\left(\hat{a}_i^\dagger\hat{a}_j+\hat{a}_j^\dagger\hat{a}_i\right),
\end{aligned}
\end{equation}
where the first term describes the modes of all qubits and couplers, the second term corresponds to the two cable modes, and the third term accounts for the interactions between nearest-neighbor modes.

\begin{table}[h]
\caption{All parameters in the full circuit Hamiltonian}\label{tab1}%
\begin{tabular}{@{}llll@{}}
\toprule
Parameters & Value  & Parameters & Value\\
\midrule
$E_{J1}/2\pi$(GHz)   & 13.5   &$E_{C1}/2\pi$(MHz)   & 221 \\
$E_{J2}/2\pi$(GHz)   & 14.35  &$E_{C2}/2\pi$(MHz)   & 221\\
$E_{J5}/2\pi$(GHz)   & 32.3   &$E_{C3}/2\pi$(MHz)   & 1.74\\
$E_{J6}/2\pi$(GHz)   & 32.4   &$E_{C4}/2\pi$(MHz)   & 1.74\\
$E_{J7}/2\pi$(GHz)   & 32.3   &$E_{C5}/2\pi$(MHz)   & 219\\
$E_{J8}/2\pi$(GHz)   & 32.4   &$E_{C6}/2\pi$(MHz)   & 218\\
$E_{J9}/2\pi$(GHz)   & 9.69   &$E_{C7}/2\pi$(MHz)   & 219\\
$E_{J10}/2\pi$(GHz)  & 9.69   &$E_{C8}/2\pi$(MHz)   & 218 \\
$E_{L3}/2\pi$(GHz)   & 1450   &$J_{15}/2\pi$(MHz)   & 283  \\
$E_{L4}/2\pi$(GHz)   & 1755   &$J_{28}/2\pi$(MHz)   & 283  \\

$J_{56}/2\pi$(MHz)   & 43     &$J_{78}/2\pi$(MHz)   & 43  \\
$J_{36}/2\pi$(MHz)   & 25     &$J_{46}/2\pi$(MHz)   & 25  \\
$J_{37}/2\pi$(MHz)   & 25     &$J_{47}/2\pi$(MHz)   & -25 \\

$\omega_\mathrm{q_1}/2\pi$(GHz)   &4.55    &$\omega_\mathrm{q_2}/2\pi$(GHz)   &4.70 \\ 
$\omega_\mathrm{cb_1}/2\pi$(GHz)  &4.44    &$\omega_\mathrm{cb_2}/2\pi$(GHz)  &4.90 \\
$\omega_{Cp_{1A}}/2\pi$(GHz)      &7.33    &$\omega_{Cp_{1B}}/2\pi$(GHz)      &7.78 \\
$\omega_{Cp_{2A}}/2\pi$(GHz)      &7.35    &$\omega_{Cp_{2B}}/2\pi$(GHz)      &7.80 \\
\botrule
\end{tabular}
\end{table}

\subsection{Effective low-energy truncation and cable mode}

Following standard canonical quantization, the phase $\hat{\varphi}_{i}$ and its conjugate variable $\hat{n}_{i}$ satisfy the commutation relation $\left[\hat{\varphi}_{i},\hat{n}_{j}\right]=i\delta_{ij}$. In the phase representation, the number operator acts as a differential operator, 
$\hat{n}_i=-i\frac{\partial}{\partial\varphi_i}$, with corresponding eigenfunctions proportional to $e^{in_i\varphi_i}$. By projecting the system onto this discrete charge basis, the operators can be expressed in the following matrix forms:
\begin{equation}\hat{n}_i=\begin{pmatrix}-N&&\\&\ddots&\\&&N\end{pmatrix},\end{equation}
\begin{equation}\cos\hat{\varphi}_{i}=\frac{1}{2}\begin{pmatrix}&1\\1&&\ddots\\&\ddots&&1\\&&1\end{pmatrix},\end{equation}
\begin{equation}\sin\hat{\varphi}_i=\frac{1}{2i}\begin{pmatrix}&-1\\1&&\ddots\\&\ddots&&-1\\&&1\end{pmatrix},\end{equation}
where we symmetrically truncate the number of Cooper pairs at $N$. For $\cos(\hat{\varphi}_i-\hat{\varphi}_j+2\pi\Phi_{\mathrm{ext}}^{(l)})$ can be expanded as $[\cos\hat{\varphi}_{i}\cos\hat{\varphi}_{j}+\sin\hat{\varphi}_{i}\sin\hat{\varphi}_{j}]\cos(2\pi\Phi_{\mathrm{ext}}^{(l)})-[\sin\hat{\varphi}_{i}\cos\hat{\varphi}_{j}-\cos\hat{\varphi}_{i}\sin\hat{\varphi}_{j}]\sin(2\pi\Phi_{\mathrm{ext}}^{(l)})$.

In our calculations, we set $N = 50$, which is sufficient to accurately capture the low-energy properties of each local mode. However, for the full circuit, a direct treatment in the complete tensor-product charge basis becomes computationally intractable due to the rapid growth of the Hilbert-space dimension with the number of degrees of freedom. To address this issue, we adopt an eigenstate-based truncation scheme. Specifically, each local Hamiltonian is first diagonalized in the truncated charge basis, after which only the lowest few eigenstates are retained to define an effective low-energy subspace. The full system Hamiltonian and relevant operators are then projected onto the tensor product of these reduced local subspaces. This approach significantly lowers the computational cost while maintaining an accurate description of the low-energy spectrum and dynamics. The numerical procedure is summarized in the Algorithm 1. To verify convergence of the low-energy spectrum over the operating range, we retain the lowest three eigenstates for each qubit, five for each DTC, and two for each cable mode to construct an effective low-energy subspace. By applying this methodology, the dimension of the global system Hamiltonian is dramatically compressed to $22500\times22500$.

\begin{figure*}
\centering
\includegraphics[width=1\linewidth]{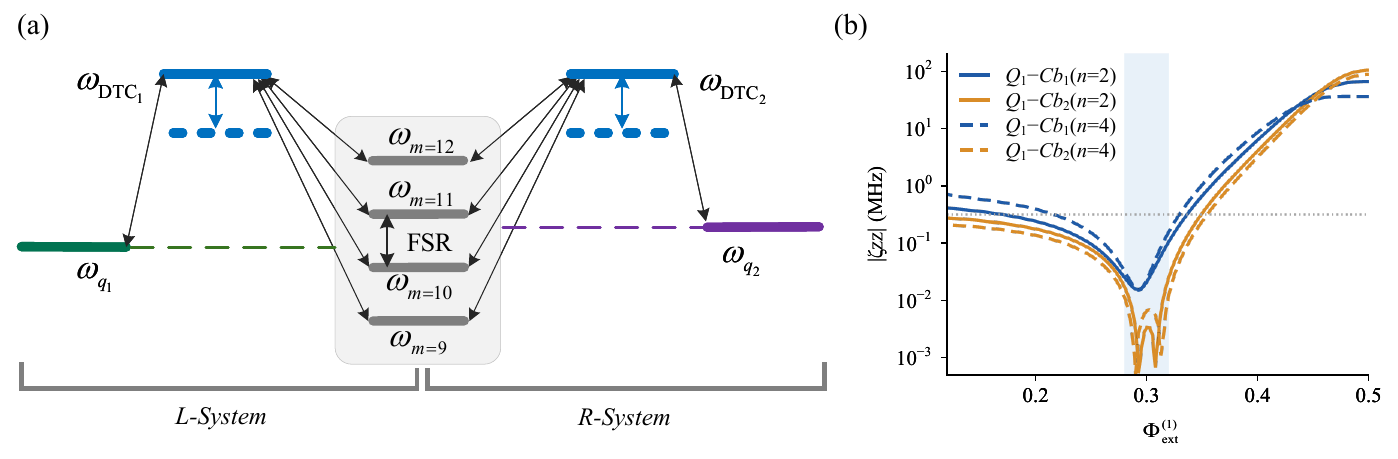}
\caption{\label{Fig5} Four-mode cable model and its impact on the effective $ZZ$ interaction. (a) Schematic energy-level diagram of the distributed architecture incorporating four adjacent longitudinal modes of the coaxial cable ($\omega_{m=9}$--$\omega_{m=12}$), separated by the free spectral range (FSR). Fixed-frequency qubits $q_{1}$ and $q_{2}$ (green and purple solid lines, respectively) are locally coupled to tunable double-transmon couplers, $\mathrm{DTC}_{1}$ and $\mathrm{DTC}_{2}$ (blue solid lines). Blue arrows and dashed lines indicate the frequency-tuning ranges of the DTCs. Each DTC couples simultaneously to its local qubit and to all retained cable modes, as indicated by black double-headed arrows. (b) Comparison of the $ZZ$ coupling strengths calculated using two-mode and four-mode cable truncations. The $ZZ$ interactions between $Q_{1}$ and the two central cable modes, $Cb_{1}$ and $Cb_{2}$, are shown as functions of the external flux $\Phi_{\mathrm{ext}}$ in the L-system.}
\end{figure*}

In the model above, we initially retained two cable modes. To assess the effect of mode truncation, we further extended the model to include four cable modes, and the corresponding energy-levels of the qubit-cable system are shown in Fig.~\ref{Fig5}(a). We then calculated the $ZZ$ coupling strengths between the qubit and the two central cable modes in the $L~system$. As shown in Fig.~\ref{Fig5}(b), the $ZZ$ couplings obtained from the four-mode model exhibit no appreciable deviation from those of the two-mode model within the parameter range relevant to our gate operation, suggesting that the dominant contribution is already captured by the central cable modes.

\begin{algorithm}[h]
\caption{Numerical construction of truncated transmon operators}
\label{alg:transmon}
\SetKwInOut{Input}{Input}
\SetKwInOut{Output}{Output}
\SetKwFunction{Diag}{diag}
\SetKwFunction{Arange}{arange}
\SetKwFunction{Ones}{ones}
\SetKwFunction{Zeros}{zeros}
\SetKwFunction{Qobj}{Qobj}
\SetKwFunction{Eigenstates}{Eigenstates}
\SetKwFunction{ConvertToCSR}{ConvertToCSR}
\SetKw{KwRet}{return}

\Input{Josephson energy $E_J$, charging energy $E_C$, charge-basis cutoff $\mathrm{Num}$, truncated dimension $N_{\mathrm{Fock}}$}
\Output{Truncated Hamiltonian $H_{\mathrm{tru}}$ and truncated charge operator $N_{\mathrm{op,tru}}$}

\emph{Step 1: Construct untruncated transmon Hamiltonian}\;
$N \leftarrow \mathrm{Num}$\;
$m \leftarrow \Diag\!\bigl(4E_C(\Arange(-N,N+1))^2\bigr)
-\frac{E_J}{2}\Bigl(\Diag(\Ones(2N),1)+\Diag(\Ones(2N),-1)\Bigr)$\;

\emph{Step 2: Define charge number operator}\;
$N_{\mathrm{op}} \leftarrow \Qobj\!\bigl(\Diag(\Arange(-N,N+1))\bigr)$\;

\emph{Step 3: Convert Hamiltonian to Qobj form and solve eigenstates}\;
$H \leftarrow \Qobj(m)$\;
$(\mathit{vals},\mathit{vecs}) \leftarrow \Eigenstates(H)$\;

\emph{Step 4: Truncate operators in the eigenbasis}\;
$H_{\mathrm{tru}} \leftarrow \Zeros(N_{\mathrm{Fock}},N_{\mathrm{Fock}})$\;
$N_{\mathrm{op,tru}} \leftarrow \Zeros(N_{\mathrm{Fock}},N_{\mathrm{Fock}})$\;

\For{$i \leftarrow 0$ \KwTo $N_{\mathrm{Fock}}-1$}{
    \For{$j \leftarrow 0$ \KwTo $N_{\mathrm{Fock}}-1$}{
        $H_{\mathrm{tru}}[i,j] \leftarrow \mathit{vecs}[i]^{\dagger} H\, \mathit{vecs}[j]$\;
        $N_{\mathrm{op,tru}}[i,j] \leftarrow \mathit{vecs}[i]^{\dagger} N_{\mathrm{op}}\, \mathit{vecs}[j]$\;
    }
}
\emph{Step 5: Convert to Qobj object and CSR sparse format}\;
$H_{\mathrm{tru}} \leftarrow \ConvertToCSR(\Qobj(H_{\mathrm{tru}}))$\;
$N_{\mathrm{op,tru}} \leftarrow \ConvertToCSR(\Qobj(N_{\mathrm{op,tru}}))$\;

\KwRet{$H_{\mathrm{tru}}, N_{\mathrm{op,tru}}$}\;

\end{algorithm}

\subsection{Optimization Procedure}
To identify pulse parameters that realize the remote CZ gate, we numerically optimize the control waveforms of the two tunable couplers using a cost function that penalizes both conditional-phase error and leakage out of the computational subspace.
For a given set of pulse parameters ($\lambda_1$, $\Phi_{f1}$) and ($\lambda_2$, $\Phi_{f2}$), the full system evolution is obtained by solving the Schrödinger equation under the time-dependent Hamiltonian, with the initial state $|\psi_{\mathrm{in}}\rangle$. The simulated Hilbert space includes the two fixed-frequency qubits, the retained cable modes, and the four coupler degrees of freedom. The final state is then obtained as
\begin{equation}|\psi_f\rangle=U(\lambda_1,\Phi_{f1};\lambda_2,\Phi_{f2})|\psi_\mathrm{in}\rangle,\end{equation}
where $U$ is the time-evolution operator generated by $H(t)$.

To characterize the entangling action of the pulse sequence, we evaluate the relative phase accumulated among the computational basis components of the final state. Let
\begin{equation}\mathcal{H}_{\mathrm{comp}}=\mathrm{span}\{\left|00\right\rangle,\left|01\right\rangle,\left|10\right\rangle,\left|11\right\rangle\}\end{equation}
be the two-qubit computational subspace. We choose the input state $|\psi_{\mathrm{in}}\rangle$ such that all four computational basis states have nonzero overlap with the evolved state, which allows the relative phases of the computational components to be extracted from a single simulation.

The effective conditional phase is defined as
\begin{equation}
\begin{aligned}
\phi_{\mathrm{cond}}=\arg\left(\langle11|\psi_f\rangle\right)-\arg\left(\langle01|\psi_f\rangle\right)\\
-\arg\left(\langle10|\psi_f\rangle\right)+\arg\left(\langle00|\psi_f\rangle\right).
\end{aligned}
\end{equation}
For an ideal CZ gate, one expects $\phi_{\mathrm{cond}}=\pi.$ We define the phase-mismatch term as
\begin{equation}\mathcal{C}_\phi=\left(\left|\arg(e^{i\phi_{\mathrm{cond}}})\right|-\pi\right)^2.\end{equation}

In addition to generating the target conditional phase, a high-quality gate must suppress population transfer to non-computational states. We therefore define the projector onto the computational subspace as
\begin{equation}P_\mathrm{comp}=|00\rangle\langle00|+|01\rangle\langle01|+|10\rangle\langle10|+|11\rangle\langle11|,\end{equation}
and quantify the leakage by
$L=1-\langle\psi_f|P_{\mathrm{comp}}|\psi_f\rangle.$
Here, $L$ measures the total population outside the computational subspace at the end of the gate.

The cost function used in the numerical calibration is defined as \begin{equation}\mathcal{C}=\mathcal{C}_\phi+|L|,\end{equation}
or explicitly,
\begin{equation}\mathcal{C}=\left(|\arg(e^{i\phi_{\mathrm{cond}}})|-\pi\right)^2+|1-\langle\psi_f|P_{\mathrm{comp}}|\psi_f\rangle|.\end{equation}
Minimizing $\mathcal{C}$ drives the pulse parameters toward a regime where the operation produces the target CZ conditional phase while maintaining the final state within the computational subspace. In the numerical implementation, for each trial set of pulse parameters $(\lambda_1,\Phi_{f1},\lambda_2,\Phi_{f2})$,  we solve the full time-dependent evolution, extract $\phi_{\mathrm{cond}}$ and $L$ from the resulting final state, and use the corresponding value of $\mathcal{C}$ as the objective for optimization.

Based on the optimization protocol described above, we systematically evaluated the performance of the CZ gate across various total gate durations, as shown in Fig. \ref{Fig3}(d). For each specific duration, we independently optimized the control parameters to minimize the cost function, extracting the total gate infidelity ($1-F$), the population leakage out of the computational subspace, and the contribution of the residual conditional phase error. Within the investigated temporal regime, the total gate error closely follows the leakage curve, indicating that the simulated coherent gate operation is predominantly leakage-limited.

To explicitly demonstrate the efficacy of our optimization protocol, we illustrate a representative convergence trajectory of the algorithm for a target gate duration of $T= 350$~ns in Fig.~\ref{Fig3}(c). During the initial iterations, the optimizer reduces the conditional-phase error, bringing the conditional phase (red line) close to the target value of $\pi$. Once the phase penalty is minimized, the total cost function (blue line) becomes entirely dominated by the population leakage (orange line). As the iterative search progresses, the algorithm effectively fine-tunes the pulse shaping parameters to suppress non-adiabatic transitions, driving both the leakage and the overall cost down until they reach the predetermined convergence threshold of $10^{-4}$. The optimal parameter set, extracted at the terminal step marked by the red star, yields a high-fidelity, leakage-limited remote CZ operation within the simulated model.

\bmhead{Acknowledgements}
This work was supported by the National Key Research and Development Program of China (Grant No. 2024YFB4504101). We thank Dr. Peng Zhao from the Quantum Science Center of the Guangdong-Hong Kong-Macao Greater Bay Area for his guidance on circuit design.

\section*{Declarations}

\begin{itemize}

\item Funding 

This work was supported by the National Key Research and Development Program of China (Grant No. 2024YFB4504101).

\item Conflict of interest/Competing interests 

The authors declare no competing interests.

\item Ethics approval and consent to participate

Not applicable
\item Consent for publication

Not applicable
\item Data availability

The data that support the findings of this study are available from the corresponding author upon reasonable request.

\item Code availability 

Codes are available from the corresponding author upon reasonable request.

\item Author contribution

Z.S. and W.W. conceived the project and supervised the research. B.Y. developed the theoretical model, performed the circuit-level numerical simulations, and analyzed the data. C.Z. and H.H. assisted with the theoretical derivations and simulation setup. Y.F., C.H., and S.W. contributed to the optimization of physical parameters and data validation. H.S., Q.M., B.Z., and F.L. participated in scientific discussions regarding hardware implementation and architectural scalability. B.Y. wrote the original manuscript with critical input from W.W. and Z.S. All authors discussed the results, reviewed the manuscript, and contributed to the final version.

\end{itemize}

\bibliography{sn-bibliography}

\end{document}